\documentclass[letterpaper]{aa}
\usepackage{graphicx}
\usepackage{natbib}
\usepackage{txfonts}

\bibpunct{(}{)}{;}{a}{}{,}

\begin{document}

   \title{Oxygen isotopic ratios in galactic clouds along the line of sight towards \object{Sagittarius~B2}
          \thanks{Based on observations with ISO, an ESA project with instruments funded by ESA Member States (especially  the PI countries: France, Germany, the Netherlands and the United Kingdom) with the participation of ISAS and NASA.}
   }

   \author{E. T. Polehampton
           \inst{1}
           \and
           J.-P. Baluteau
           \inst{2}
           \and
           B. M. Swinyard
           \inst{3}
            }

   \institute{Max-Planck-Institut f\"{u}r Radioastronomie, Auf dem H\"{u}gel 69, 53121 Bonn, Germany \\
     \email{epoleham@mpifr-bonn.mpg.de}
   \and
   Laboratoire d'Astrophysique de Marseille, CNRS \& Universit\'e de Provence, BP 8, F-13376 Marseille Cedex 12, France\\
     \email{Jean-Paul.Baluteau@oamp.fr}
   \and
   Rutherford Appleton Laboratory, Chilton, Didcot, Oxfordshire, OX11 0QX, UK\\
     \email{B.M.Swinyard@rl.ac.uk} }

   \date{Received  / accepted }

 \abstract{

As an independent check on previous measurements of the isotopic abundance of oxygen through the Galaxy, we present a detailed analysis of the ground state rotational lines of $^{16}$OH and $^{18}$OH in absorption towards the giant molecular cloud complex, Sagittarius B2. We have modelled the line shapes to separate the contribution of several galactic clouds along the line of sight and calculate $^{16}$OH/$^{18}$OH ratios for each of these features. The best fitting values are in the range 320--540, consistent with the previous measurements in the Galactic Disk but slightly higher than the standard ratio in the Galactic Centre. They do not show clear evidence for a gradient in the isotopic ratio with galactocentric distance. The individual $^{16}$OH column densities relative to water give ratios of [H$_{2}$O/OH]=0.6--1.2, similar in magnitude to galactic clouds in the sight lines towards \object{W51} and \object{W49}. A comparison with CH indicates [OH/CH] ratios higher than has been previously observed in diffuse clouds. We estimate OH abundances of 10$^{-7}$--10$^{-6}$ in the line of sight features.

                 }

   \titlerunning{Oxygen isotopic ratios towards \object{Sgr~B2}}
   \maketitle
%

\section{Introduction}

The OH radical is one of the key oxygen bearing species in the interstellar medium (ISM). Not only is it a good diagnostic of physical and chemical conditions \citep[e.g.][]{goicoechea_e}, but in addition, it provides a very good way to investigate the relative abundance of oxygen isotopes via its isotopologues $^{16}$OH, $^{18}$OH and $^{17}$OH. Chemical fractionation reactions that might distort the oxygen isotopic ratios in molecular species are not thought to be important \citep{langer} and this means that observations of OH can be used directly to determine the ratios $^{16}$O/$^{18}$O and $^{18}$O/$^{17}$O. 

These values are important as they are set by stellar processing and outflow mechanisms and so constrain models of galactic chemical evolution \citep[e.g.][]{prantzos_b}. $^{16}$O is a primary product of stellar nucleosynthesis, produced directly from the primordial elements H and He \citep[see][]{wilsonb}. Both $^{17}$O and $^{18}$O are secondary products which require heavier elements from previous nuclear burning for their production. Chemical evolution models \citep[e.g.][]{prantzos_b} show that the primary/secondary ratios, $^{16}$O/$^{17}$O and $^{16}$O/$^{18}$O, should fall with decreasing galactocentric distance due to the increased processing rate towards the Galactic Centre. These ratios should also fall with time due to the build up of the secondary elements in the ISM. 

Previous measurements of $^{16}$O/$^{18}$O in the ISM have mainly been made via the radio lines of H$_{2}$CO and CO towards molecular clouds \citep[see][]{wilson}. However, these lines are generally optically thick in the most abundant isotopologues and so double ratios such as H$_{2}^{13}$C$^{16}$O/H$_{2}^{12}$C$^{18}$O are used. This relies on accurate knowledge of $^{12}$C/$^{13}$C which \emph{is} subject to chemical fractionation in molecular species \citep[e.g.][]{langer}. 

A very good way to get around these difficulties is to use the OH molecule - several measurements using its 18~cm $\Lambda$-doubling transitions have been made towards the Galactic Centre \citep{whiteoak_b,williams}, although in some circumstances these can be complicated by excitation effects within the hyperfine levels \citep{bujarrabal}. This problem can be avoided by using the far-infrared (FIR) rotational lines which should not be affected by hyperfine excitation anomalies. These provide an excellent way to independently check the ratios determined at radio wavelengths.

Rotational transitions of OH have previously been studied towards the Orion Kleinmann-Low nebula using the Kuiper Airborne Observatory \citep[][ and references therein]{melnick} where both $^{16}$OH and $^{18}$OH were used to constrain the physical conditions of the source.  More recently, \citet{goicoechea_b} have used the Infrared Space Observatory (ISO) satellite to observe OH in the envelope of the giant molecular cloud complex Sagittarius~B2 (\object{Sgr~B2}). The lowest energy transitions show absorption due to both $^{16}$OH and $^{18}$OH in \object{Sgr~B2} itself and in foreground features intersecting the line of sight. They indicate that the isotopic ratios are broadly similar to the previous radio results.

\begin{figure}
\resizebox{\hsize}{!}{\includegraphics{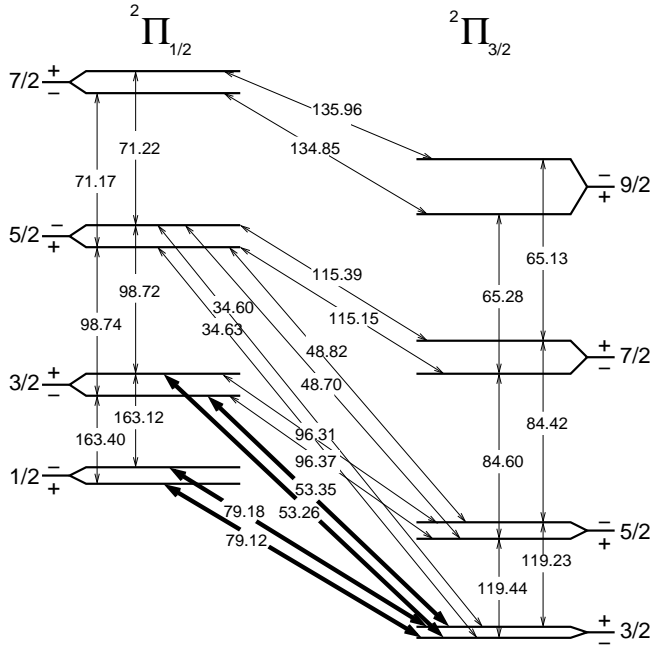}}
\caption {Low lying rotational energy levels in the electronic and vibrational ground state of the $^{16}$OH molecule, adapted from \citet{brown}. The $\Lambda$-doubling of each level is shown but the splitting has been exaggerated for clarity. The relative spacing of the levels in energy is only approximate. The $J$ values for each level are shown and the wavelength of each transition is given in $\mu$m. The transitions used in the analysis presented here are shown as thick arrows.}
\label{oh_levels}
\end{figure}

\begin{figure}
\resizebox{\hsize}{!}{\includegraphics{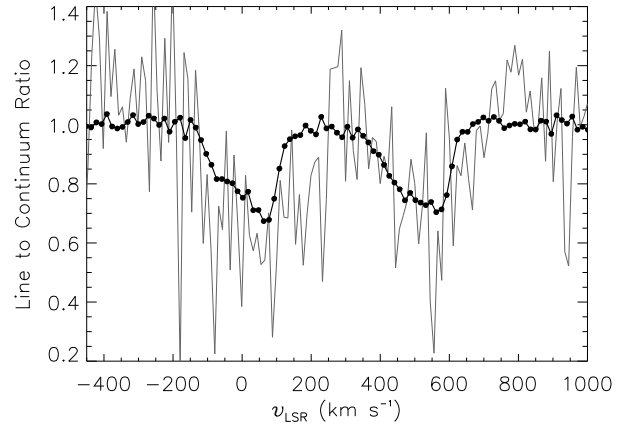}}
\caption {The $^{16}$OH $^{2}\Pi_{1/2}$--$^{2}\Pi_{3/2}$ $J$=3/2--3/2 transition at 53.26~$\mu$m from non-prime data (black) and prime data (grey). Both datasets are binned at 1/4 of their respective resolution elements (non-prime: 61~km~s$^{-1}$ and prime: 45~km~s$^{-1}$). The gain in signal-to-noise in the non-prime data is a factor of $\sim$9.}
\label{53fpsfpl}
\end{figure}

In this paper we conduct a more detailed investigation into ISO observations towards \object{Sgr~B2}, with the aim of separating the relative abundances of $^{16}$OH and $^{18}$OH in individual absorption components in the line of sight. A similar comparison of $^{18}$OH with $^{17}$OH has already been presented by \citet{polehampton_c}. We have used data from a wide spectral survey carried out with the ISO Long Wavelength Spectrometer \citep[LWS;][]{clegg} Fabry-P\'{e}rot mode. A combination of prime and non-prime data from the survey allowed us to increase the signal-to-noise ratio over the standard data and derive an accurate and consistent calibration for two ground state transitions of $^{16}$OH and one transition of $^{18}$OH. 

After presenting the observations and results, we describe a model of the line shape to separate the line of sight absorption into 10 velocity components (Sect.~\ref{modelling}). In Sect.~\ref{ratios} we assign the $^{16}$OH/$^{18}$OH ratio for each component to a galactocentric distance and compare with previous determinations of the isotopic abundances. We discuss the final $^{16}$OH column densities in relation to other related species in Sect.~\ref{otherspecies}.

\section{Observations and data reduction \label{data_red}}

\object{Sgr~B2} was observed as part of a wide spectral survey using the ISO LWS Fabry-P\'{e}rot (FP) mode L03. Unbiased coverage of the whole LWS spectral range (47--196~$\mu$m) was carried out using 36 separate observations with a spectral resolution of $30$--$40$~km~s$^{-1}$. No other object outside of the Solar System was observed over the complete LWS spectral range in this way. Results from this survey have been presented by \citet{ceccarelli, polehampton_b, vastel_b, polehampton_c, polehampton_ch}. \object{Sgr~B2} was also extensively observed by the LWS FP in narrow wavelength scans using the L04 mode \citep[see][]{goicoechea_d}. We have used several of these L04 observations to improve the signal-to-noise ratio in the L03 data for $^{18}$OH. The ISO TDT numbers for all the observations used in this paper are detailed in Table~\ref{tdts}.

The LWS beam had an effective diameter of approximately 80$\arcsec$ \citep{gry} and L03 observations were pointed at coordinates $\alpha=17^{\mathrm{h}}47^{\mathrm{m}}21.75^{\mathrm{s}}$, $\delta=-28\degr 23\arcmin 14.1\arcsec$ (J2000). This gave the beam centre an offset of 21.5$\arcsec$ from the main FIR peak at \object{Sgr~B2~(M)} - this pointing was used to exclude the source \object{Sgr~B2~(N)} from the beam. The additional L04 observations were pointed directly towards the nominal position of \object{Sgr~B2~(M)}, but careful comparison showed no significant difference to the L03 observations. Each observation had a spectral sampling interval of 1/4 resolution element with each point repeated 3--4 times in L03 mode and 11--15 times in L04 mode.

\begin{figure*}
\centering
\resizebox{\hsize}{!}{\includegraphics{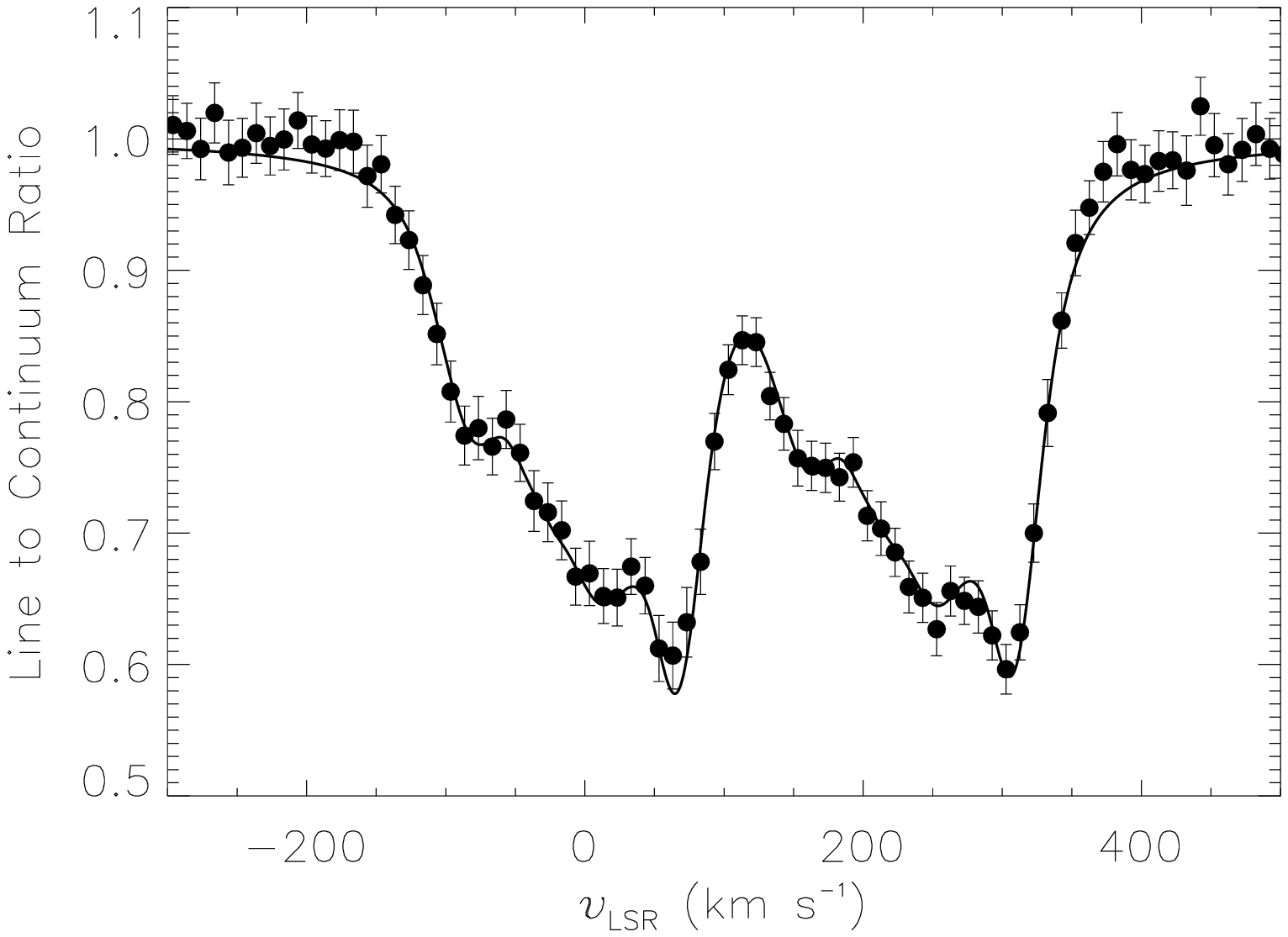}\includegraphics{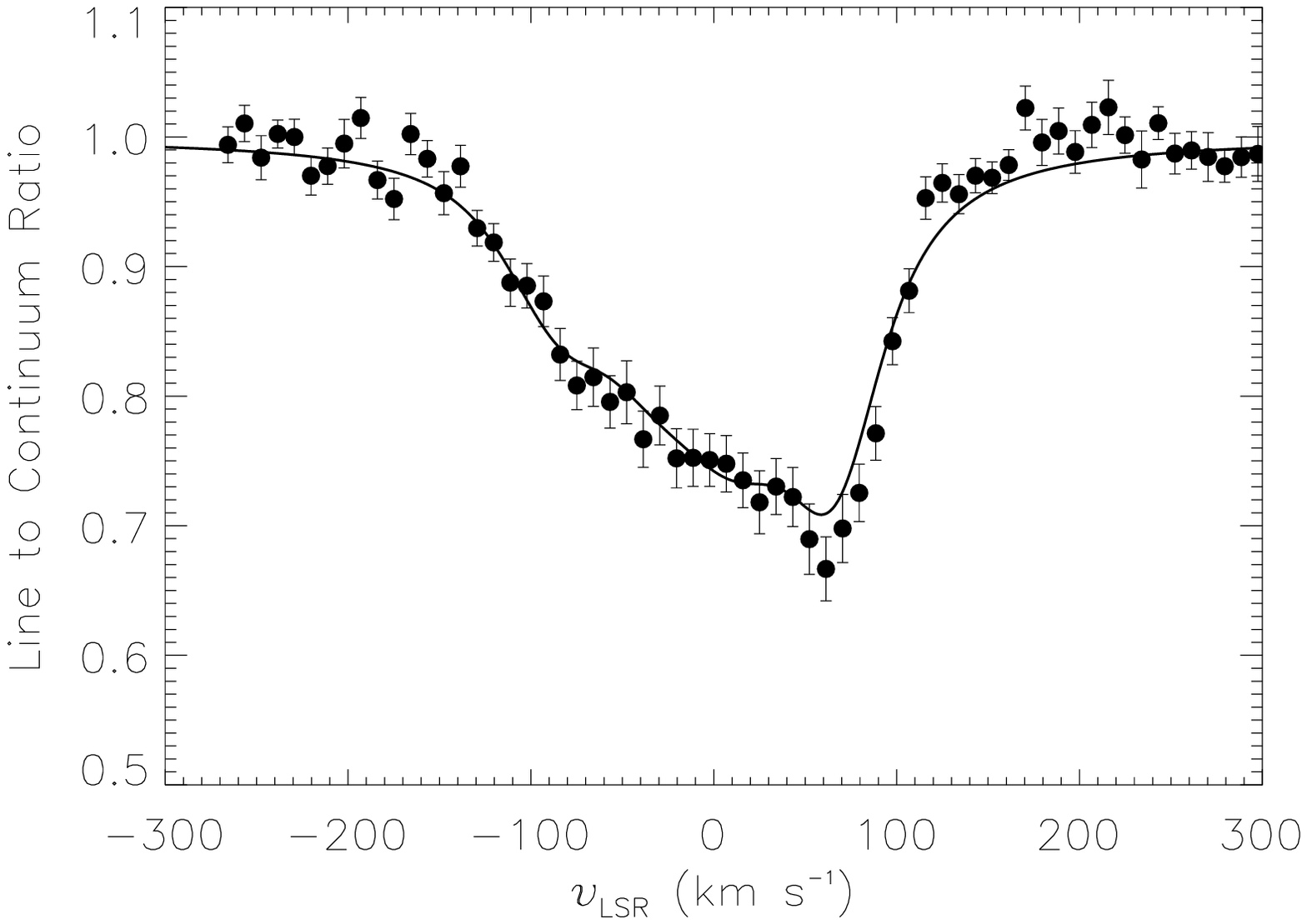}}
\caption{Data for the $^{16}$OH cross ladder ($^{2}\Pi_{1/2}$--$^{2}\Pi_{3/2}$) transitions, $J$=1/2--3/2 at 79~$\mu$m (left; binned at 1/4 resolution element) and $J$=3/2--3/2 at 53~$\mu$m (right; binned at $\sim$1/6 resolution element), plotted as line to continuum ratio. The errors shown represent the combined statistical and systematic uncertainty in each bin. The two $\Lambda$-doublet components have been averaged for the 53~$\mu$m line. The best fitting model accounting for the absorbing clouds in the line of sight is shown as the solid line (see text).}
\label{5379}
\end{figure*}

In addition to the primary spectral survey data, each line was generally included in at least one other L03 observation. This occurred because the LWS used 10 detectors, which always recorded data in their own spectral ranges (the prime observations consist of data from a single detector for which instrument settings were optimised). We have included these 'non-prime' data to further improve the signal-to-noise ratio. 

Each observation was carefully reduced using the LWS offline pipeline (OLP) version 8 (for FP data the difference between OLP version 8 and the latest version 10 is not significant). Further processing was then carried out interactively using routines that appeared in the LWS Interactive Analysis package version 10 \citep[LIA10:][]{lim_d} and the ISO Spectral Analysis Package \citep[ISAP:][]{sturm}. The method  included determination of accurate dark currents (including stray light), interactive division of each mini-scan by the LWS grating response profiles and careful scan by scan removal of glitches \citep[see][]{polehampton_b,polehampton_c}.

Before co-adding the data for each line, the wavelength scale of each observation was corrected to the local standard of rest (LSR). Non-prime observations were carefully checked against the equivalent prime measurement to align the line centres. This is particularly important for non-prime data measured with the LWS long wavelength FP (FPL) below 70~$\mu$m (outside its nominal operating range) because the standard FPL wavelength calibration did not include short wavelength data \citep{gry}. For the $^{16}$OH $J$=3/2--3/2 cross ladder transition at 53~$\mu$m (Fig.~\ref{oh_levels} shows the energy levels for OH), we have used only non-prime FPL data. This is because it has a significantly higher signal-to-noise ratio than the prime data \citep[due to the higher transmission of FPL compared to the short wavelength FP, FPS;][]{polehampton_d}. The absolute wavelength alignment was difficult to determine purely by comparison with the noisy prime data and so a velocity shift was allowed as a free parameter in the modelling (see Sect.~\ref{model}). Figure~\ref{53fpsfpl} shows the resulting shift cross-checked with the prime data. The gain in signal-to-noise achieved by using FPL rather than FPS is approximately a factor of 9. The only cost in using these data is a reduction in spectral resolution from 45~km~s$^{-1}$ (for FPS) to 61~km~s$^{-1}$ (for FPL).

After co-addition, the continuum around each line was fitted with a 3rd order polynomial baseline which was then divided into the data to obtain the relative depth of the lines below the continuum. This effectively bypassed the large systematic uncertainties in the multiplicative calibration factors needed to obtain the absolute flux scale \citep[see][]{swinyard_b}. The remaining errors are due to detector noise, uncertainty in the dark current and the polynomial fit. 

At the resolution of the LWS, two components are visible for each OH transition due to the $\Lambda$-doublet type splitting of each rotational level. However, further hyperfine splitting is not resolved. The two $\Lambda$-doublet components showed good agreement in the data, and so where they were well separated (the $J$=3/2--3/2 cross ladder transition at 53~$\mu$m for $^{16}$OH and the $J$=5/2--3/2 transition at 120~$\mu$m for $^{18}$OH), they were co-added to further increase the signal-to-noise ratio.

\section{Results}

Figure~\ref{oh_levels} shows the low-lying rotational transitions of $^{16}$OH, most of which are included in the spectral survey range. The strongest lines observed in the survey were due to the $^{2}\Pi_{3/2}~J$=5/2--3/2 transition from the ground state at 119~$\mu$m. These lines show almost complete absorption of the FIR continuum in the range $-150$ to $+100$~km~s$^{-1}$ \citep[see][]{goicoechea_b}. Due to the large depth of the lines, the shape was strongly affected by the transient response of the LWS detectors. These detector memory effects \citep[see][]{lloyd_a} meant that successive repeated scans underestimated the depth and the continuum level following each line. Therefore, the lines were not used in the analysis presented here. There are two remaining transitions from the ground rotational state (see Fig.~\ref{oh_levels}) that occur between the $^{2}\Pi_{3/2}$ and $^{2}\Pi_{1/2}$ ladders: $J$=1/2--3/2 at 79~$\mu$m and $J$=3/2--3/2 at 53~$\mu$m. These lines are also broad but with much lower optical depth. They are shown in Fig.~\ref{5379}. The broad absorption is due to features between the Sun and Galactic Centre associated with galactic spiral arms that cross the line of sight \citep[e.g.][]{greaves94}.

Several $^{16}$OH transitions between higher energy rotational levels are observed in the survey at the velocity of \object{Sgr~B2} itself but do not occur in the line of sight clouds. These lines originate in the envelope of \object{Sgr~B2} and have been modelled by \citet{goicoechea_b}. They show that the excited OH originates in clumpy photodissociation regions (PDRs) on the edge of the \object{Sgr~B2} complex at temperatures of 40--600~K. In this paper, we concentrate on the broad absorption observed in the ground state lines at 79 and 53~$\mu$m.

\begin{figure}
\resizebox{\hsize}{!}{\includegraphics{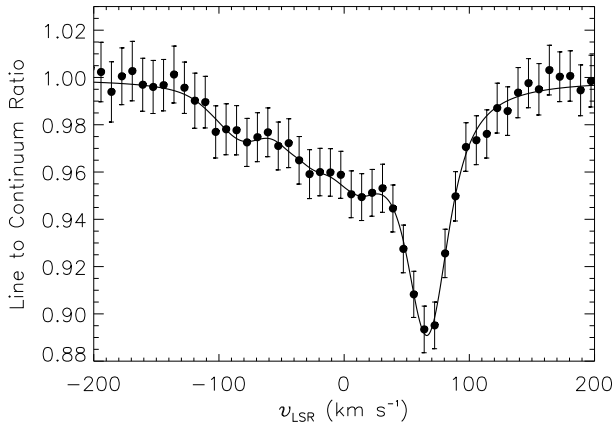}}
\caption{Co-added data from the two $\Lambda$-doublet components of the $^{18}$OH $^{2}\Pi_{3/2}~J=5/2$--$3/2$ transition. The data were binned at 1/4 resolution element and the errors represent the combined statistical and systematic uncertainty in each bin. The best fitting model accounting for absorbing clouds in the line of sight is shown (see text).}
\label{18ohfit}
\end{figure}

Due to the lower abundance of $^{18}$OH, the fundamental ground state $^{2}\Pi_{3/2}~J$=5/2--3/2 transition is much weaker and the cross-ladder and higher energy lines are not detected. Figure~\ref{18ohfit} shows the $J$=5/2--3/2 line after co-adding the two resolved $\Lambda$-doublet components. These observations were previously presented by \citet{polehampton_c}. Here we have used additional data from the L04 mode, which showed good agreement with the L03 data.

\section{Modelling of OH in the line of sight \label{modelling}}

\subsection{High resolution model \label{model}}

At the spectral resolution of the LWS, the line of sight components are blended together into a single broad absorption. We have modelled the line shape using higher spectral resolution measurements to fix the velocities and widths of each component. Comparison of high resolution spectra tracing molecular and atomic species in these clouds show very similar line widths and velocities (CO; e.g. \citet{vastel_b}, H$_{2}$CO; e.g. \citet{mehringerb}, HI; e.g. \citet{garwood}). These tracers also show a very similar velocity structure to observations of the $\Lambda$-doubling lines of OH at cm wavelengths \citep{mcgee,whiteoak_c,whiteoak_b,williams,bujarrabal}. 

We have assumed that the \ion{H}{i} measurements are a representative tracer and based our model on the published parameters from \citet{garwood} (after correcting an error in their Table~2: 11~km~s$^{-1}$ should read 1.1~km~s$^{-1}$). This method has already been used to successfully fit the absorption of CH and CH$_{2}$ \citep{polehampton_ch} and is similar to the method used to fit \ion{O}{i} and \ion{C}{ii} lines by \citet{vastel_b}.

The \ion{H}{i} observations were made using the VLA pointed close to \object{Sgr~B2~(M)}. In our model, the \ion{H}{i} data were used to fix the velocity and line width of 10 line of sight features with each component assumed to have a Gaussian line shape. The optical depth at each velocity, $\tau$, was then adjusted as a free parameter and the line-to-continuum ratio calculated from,
\begin{equation}
\label{eqn_abs}
I = I_c \exp{(-\tau)}
\end{equation}
where $I_c$ is the intensity of the continuum. The resulting spectrum was then convolved to the resolution of the LWS. In order to put the best constraint on the line shape, the two $^{16}$OH lines were fitted simultaneously with the $^{18}$OH line. Optical depths were varied in the model for the 79~$\mu$m line and then used to calculate the shape of 53~$\mu$m line assuming that both lines trace the same column density. This must be true as they originate in the same lower level. Column densities were calculated assuming a Doppler line profile with Maxwellian velocity distribution \citep[e.g.][]{spitzer},
\begin{equation}
\label{eqn_colden}
N_j= \frac{8\pi\sqrt{\pi}}{2\sqrt{\ln2}}~10^{17}~\frac{\tau_{0}~\Delta{v}} {A_{i\,j}~\lambda_{i\,j}^{3}~g_{i}/g_{j}}
\end{equation}
where $N_j$ is the column density in the lower level in cm$^{-2}$, $\tau_{0}$ is the optical depth at line centre, $\Delta{v}$ is the line width in km~s$^{-1}$, $A_{ij}$ is the Einstein coefficient for spontaneous emission in s$^{-1}$, $\lambda_{ij}$ is the wavelength in $\mu$m and $g_{i}$ is the statistical weight of state $i$. The line wavelengths and Einstein coefficients used (averaged over the unresolved hyperfine structure) are shown in Table~\ref{wavelengths}.

Further free parameters were added to the fit describing the $^{16}$OH/$^{18}$OH column density ratio, allowing the shape of the 120~$\mu$m $^{18}$OH line to be calculated. Clouds which are thought to reside at similar galactocentric distances were forced to have identical isotope ratios - this affects the features at $-108/-82$~km~s$^{-1}$, $-52/-44$~km~s$^{-1}$ and $+53/+67$~km~s$^{-1}$. In order to account for any drift in the wavelength calibration between the lines, a free velocity shift was allowed for each line, giving a total of 20 free parameters in the fit. The most important factor constraining the velocity shift was the relatively deep and narrow absorption at the velocity of \object{Sgr~B2} in the $^{18}$OH line. The Numerical Recipes multi-parameter fitting routine, `amoeba' \citep{press}, was used to minimise $\chi^{2}$.

\begin{table*}
\caption{Column densities and $^{16}$OH/$^{18}$OH ratios for the best fit carried out to the $^{16}$OH and $^{18}$OH ground state lines. Columns 1 and 2 show the Gaussian fit parameters from the \ion{H}{i} observations of \citet{garwood}.}
\label{tab_col}
\begin{center}
\begin{tabular}{cccc}
\hline 
\hline 
Velocity       & FWHM         & $N(\mathrm{^{16}OH})$ & $^{16}$OH/$^{18}$OH \\
(km~s$^{-1}$)  & (km~s$^{-1}$)& ($10^{15}$~cm$^{-2}$) &  \\
\hline 
$-108\pm2$                    & $7\pm4$            & $0.56\pm0.39$         & $460^{+90}_{-50}$\\   
$-82\pm4$                     & $28\pm12$          & $9.0\pm3.9$           & $460^{+90}_{-50}$\\ 
$-44.0\pm0.6/(-52\pm6)^a$     & $8\pm2$ ($17\pm5$) & $3.5\pm1.8$           & $450^{+100}_{-60}$\\ 
$-24.4\pm0.8$                 & $14\pm2$           & $5.4\pm1.8$           & $370^{+110}_{-50}$\\ 
$+1.1\pm0.4$                  & $19\pm1$           & $7.6\pm3.4$           & $430^{+80}_{-40}$\\
$+15.7\pm0.3$                 & $7\pm1$            & $3.6\pm2.3$           & $360^{+80}_{-50}$\\
$+31\pm4$                     & $21\pm14$          & $5.4\pm3.8$           & $540^{+100}_{-50}$\\
$+66.7\pm0.5/(+52.8\pm0.9)^b$ & $16\pm1$ ($11\pm2$)& $32^{+6}_{-4}$        & $320^{+70}_{-30}$\\ 
\hline
Total       &         & 67.0                  &       \\
\hline 
\end{tabular}
\end{center}
$^a$ These two components are too close to separate in our model and the best fit line shape
requires only the $-44.0$~km~s$^{-1}$ component.\\
$^b$ The \ion{H}{i} data resolve 2 components in \object{Sgr~B2}. However, our fit requires
only one of these (at 66.7~km~s$^{-1}$) to reproduce the ISO spectrum.\\
\end{table*}

In order to ensure that the algorithm did not converge at a false minimum, the initial conditions were set close to their best values and the routine was restarted at the first convergence point. Solutions with negative absorption were avoided by setting $\chi^{2}$ to be high when the optical depth went below zero. 

The best fitting model is shown plotted with the data for each line in Figs.~\ref{5379}~and~\ref{18ohfit}. The final $^{16}$OH column densities and $^{16}$OH/$^{18}$OH ratios are shown in Table~\ref{tab_col}. In the line of sight clouds where no absorption is observed from higher energy levels \citep{goicoechea_b}, the ground state population is a good measure of the total OH column density. The fitted components are all optically thin except at the velocity of \object{Sgr~B2} itself, where optical depths of 2.5--3.3 were found for the $^{16}$OH lines. However, these are not high enough to break the assumption of purely Doppler line profiles used in the model and calculation of column densities.

The best fitting velocity shift applied to the $^{16}$OH $J$=1/2--3/2 line at 79~$\mu$m was 8.8~km~s$^{-1}$ and for the $^{18}$OH $J$=5/2--3/2 line at 120~$\mu$m was $-1.9$~km~s$^{-1}$. These shifts centred the deepest absorption on the \ion{H}{i} component at $+67$~km~s$^{-1}$ and are within the uncertainty in FP wavelength calibration \citep[$<11$~km~s$^{-1}$;][]{gry}. However, for the $^{16}$OH $J$=3/2--3/2 line at 53~$\mu$m using data taken with non-prime detectors, a larger shift of 28~km~s$^{-1}$ was required. This shift is due to the fact that this line was observed with FPL outside of its nominal range (see Sect.~\ref{data_red}). In order to determine if this large velocity offset was reasonable, we compared the shifted data to the noisier prime observation \citep[in which the accuracy of the wavelength calibration should be better than 6~km~s$^{-1}$;][]{gry}. The two line components show good agreement (see Fig.~\ref{53fpsfpl}) and the line shift is consistent with that found for other spectral lines observed using FPL below 70~$\mu$m \citep{polehampton_thesis}.

The final fit shows that only 8 of the 10 velocity components are necessary to reproduce the observed line shape. At the velocity associated with \object{Sgr~B2}, the width of the absorption in the $^{18}$OH line is too narrow to allow strong absorption by both the $+53$ and $+67$~km~s$^{-1}$ components observed for \ion{H}{i}. This is consistent with the results obtained by fitting the ground state line of CH \citep{polehampton_ch}. Also, the two components at $-52$ and $-44$~km~s$^{-1}$ are too closely spaced to be separated in our fit and significant optical depth was found only at $-44$~km~s$^{-1}$ (in Table~\ref{tab_col}, only one value centred at $-44$~km~s$^{-1}$ is given for the two components). 

We have also investigated the effect of further reducing the number of fit components in order to increase confidence in the final column densities and ratios. The minimum number of components that can reasonably be fitted to the spectrum is 4 \citep[e.g. following][]{neufeld_b}. We re-ran our model using 4 velocity components corresponding to the ranges used by Neufeld et al. and widths estimated from their H$_2^{18}$O spectrum observed with the Submillimeter Wave Astronomy Satellite (SWAS). These 4 empirically determined components can broadly reproduce the observed line shapes. The fit gives $^{16}$OH column densities and $^{16}$OH/$^{18}$OH ratios that are within the errors of the results from Table~\ref{tab_col} summed in the relevant velocity ranges. The $^{16}$OH/$^{18}$OH ratios obtained are, 480 (centred at $-80$~km~s$^{-1}$), 470 (centred at $-40$~km~s$^{-1}$), 510 (centred at $+10$~km~s$^{-1}$) and 290 (centred at $+65$~km~s$^{-1}$). However, we have used the high resolution \ion{H}{i} observations as a basis for our model because this can not only fully describe the line shapes but also allows the contribution of clouds at different galactocentric distances along the line of sight to be disentangled.

\subsection{Errors on fitted parameters \label{errs}}

The number of closely spaced velocity components in the fit made the modelling difficult as their separation was less than the resolution of the LWS. This meant that variation in one component could be compensated for by changing another, resulting in a relatively large uncertainty in each fitted optical depth. In addition, the results for each component are likely to be an average over several closely spaced narrow features such as those observed showing CS absorption with velocity widths $\sim$1~km~s$^{-1}$ \citep{greaves94}. However, this would only alter the total column densities if a few of the narrow features had very much higher optical depths than the others, and this does not appear to be the case in the CS data.

In order to estimate the uncertainty on each fitted parameter we performed a Monte-Carlo error analysis. The best fitting model determined by minimising $\chi^2$ was used to generate a set of synthetic spectra where each point had a mean value equal to the best fit and standard deviation equal to the original data error. We re-fitted each synthetic spectrum using the original fitting method and analysed the resulting dataset for each parameter. The results show that the errors on neighbouring optical depths are strongly correlated, but that these are not correlated with the uncertainty in $^{16}$OH/$^{18}$OH ratios (except for the \object{Sgr~B2} component). This is due to the fact that the best fit isotopologue ratios depend more on the overall line shape than on the relative optical depth in neighbouring components. The final uncertainty in column density is shown in Table~\ref{tab_col} as a combination of 1$\sigma$ errors in \ion{H}{i} line width \citep[as quoted by][]{garwood} and the modelling errors in optical depth. The errors in $^{16}$OH/$^{18}$OH ratio were determined directly from the Monte-Carlo analysis.

\subsection{Comparison with previous results}

The column density at the velocity of \object{Sgr~B2} has previously been calculated by \citet{goicoechea_b} using ISO FP data observed in the L04 mode. They used a radiative transfer model and accounted for the populations in higher energy levels up to 420~K above ground to determine $N(^{16}\mathrm{OH})=(1.5$--$2.5)\times10^{16}$~cm$^{-2}$. Lower resolution observations with the LWS grating mode show that OH is widespread across the whole \object{Sgr~B2} region with column densities in the range (2--5)$\times10^{16}$~cm$^{-2}$ \citep{goicoechea_d}. Our value associated with \object{Sgr~B2} from Table~\ref{tab_col} gives $N(^{16}\mathrm{OH})=(3.2^{+0.6}_{-0.4})\times10^{16}$~cm$^{-2}$ in the ground state level. This is slightly higher than the previous FP result but fits into the general picture reasonably well. We also find a slightly higher value than \citet{goicoechea_b} for $N(^{18}\mathrm{OH})$: $(9.8\pm1.6)\times10^{13}$~cm$^{-2}$ compared to $(6\pm2)\times10^{13}$~cm$^{-2}$. 

The column density of $^{16}$OH has also been measured from  observations of its $\Lambda$-doubling lines at 18~cm. \citet{bieging} used the interferometer of the Owens Valley Radio Observatory (OVRO) to calculate an integrated column density over the velocity range 40--89~km~s$^{-1}$ equal to $1.2\times10^{17}$~cm$^{-2}$. This would indicate that the FIR lines underestimate the column density by a factor of 4. However, if the excitation temperature of the 18~cm lines is less than the assumed 20~K \citep[see the discussion in][]{stacey}, the radio column density could be an over-estimate. Also, the 18~cm lines may sample a different component of the envelope as the depth at which the continuum is emitted occurs deeper into the cloud than in the FIR region, and the OVRO synthesised beam ($3.25\arcmin$) is larger than that of ISO. \citet{bieging} estimates that the column density in the negative velocity features is smaller than the positive velocity values by about a factor of four.

The total column density of $^{18}$OH over the whole line of sight in our fit was $(1.8\pm0.2)\times10^{14}$~cm$^{-2}$ and this compares favourably with previous observations using the Kuiper Airborne Observatory: \citet{lugten} found a total column density of $N(^{18}\mathrm{OH})\geq2\times10^{14}$~cm$^{-2}$. 

\begin{figure}
\resizebox{\hsize}{!}{\includegraphics{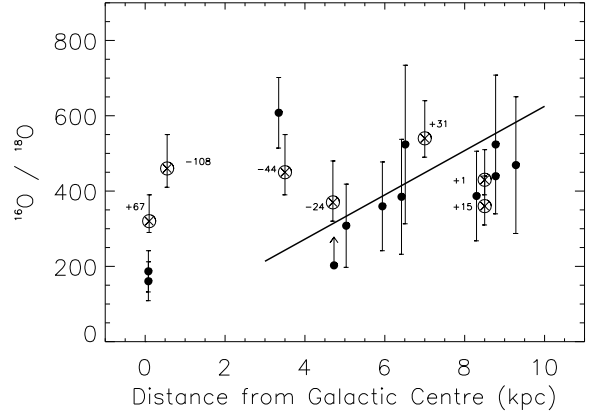}}
\caption{Values of $^{16}$OH/$^{18}$OH from Table~\ref{tab_col} plotted against distance from Galactic Centre (large circles with cross). The velocity of each component is indicated in km~s$^{-1}$ next to each point. The compilation of values from \citet{kahane} is plotted for comparison, as well as the best fitting galactic gradient found by \citet{wilson}.}
\label{fig_16-18}
\end{figure}

\section{$^{16}$OH/$^{18}$OH ratio \label{ratios}}

Table~\ref{tab_col} gives the best fitting $^{16}$OH/$^{18}$OH ratios. In order to compare with previous measurements of the isotopic ratios, a galactocentric distance must be assigned to each feature. Several of the velocity components have been well established to come from galactic spiral arm and Galactic Centre features such as the Galactic Bar at large negative velocities \citep{scoville}, the 3~kpc expanding arm at $-44$~km~s$^{-1}$ \citep[e.g.][]{burke}, the 4-5~kpc arm at $-24$~km~s$^{-1}$ \citep{menon} and \object{Sgr~B2~(M)} itself at $+50$--70~km~s$^{-1}$ \citep[e.g.][]{martin-pintado_a}. Associations for the remaining features have been proposed by \citet{greaves94}. In this case, the 1, 16 and 31~km~s$^{-1}$ features would be associated with local spiral arms. However, it has been suggested that there may be a contribution at 0~km~s$^{-1}$ by gas associated with \object{Sgr~B2} \citep[e.g][]{gardner88}.

Figure~\ref{fig_16-18} shows our results plotted against their distances from the Galactic Centre taken from \citet{greaves94}. Compilations of many observations of H$_{2}$CO through the Galaxy indicate that there is a gradient in the $^{16}$O/$^{18}$O ratio with decreasing values towards the Galactic Centre \citep{tosi,wilson,kahane}. The best fit gradient found by \citet{wilson} is overplotted in Fig.~\ref{fig_16-18} as well as the compilation of H$_{2}$CO and CO results from \citet{kahane}.

The three most uncertain distances are for the local components. At 31~km~s$^{-1}$ we find a relatively high value of $^{16}$OH/$^{18}$OH, consistent with its location close to the Sun. At 1~km~s$^{-1}$, our value is in good agreement with that found from OH $\Lambda$-doublet lines of 440 \citep{bujarrabal}. However, both the 1 and 16~km~s$^{-1}$ components have rather lower values than might be expected for local gas. The gradient of \citet{wilson} predicts a local value of $560\pm25$, although it has been measured to be even higher using HCO$^{+}$ absorption by \citet{lucas}, giving $672\pm110$. This may indicate that the local absorbing gas towards \object{Sgr~B2} has atypical isotopic abundances. An anomalously low $^{12}$C/$^{13}$C ratio has also been found in this gas by \citet{greaves95}. She rules out an error in the distance estimate and attributes the extra $^{13}$C to enrichment by stellar ejecta containing secondary nucleosynthesis products. In this case it would be natural to also find a higher abundance of $^{18}$OH. 

In comparison, the standard Solar System abundance (derived from Standard Mean Ocean Water) is ($^{16}$O/$^{18}$O)$_{\mathrm{SMOW}}=498.7\pm0.1$ \citep{baertschi}. The value in the outer layers of the Sun has been measured to be $^{16}$O/$^{18}\mathrm{O}=440\pm50$ \citep{harris} and in the Solar wind, \citet{collier} found $^{16}$O/$^{18}\mathrm{O}=450\pm130$. This should represent the conditions when the Sun was formed and models predict that it should be higher than the current ratio in the local ISM due to the build up of secondary elements from stellar processing \citep[e.g.][]{prantzos_b}. In order to explain the low Solar System abundances compared to the ISM, it has been proposed that the Sun was formed closer to the Galactic Centre \citep{wielen}.

In the Galactic Centre, the isotope ratio is generally taken to be 250 \citep{wilson}. This low value has been derived from measurements of the $\Lambda$-doublet lines of $^{16}$OH and $^{18}$OH \citep[e.g.][]{williams, whiteoak_b}. However, \citet{bujarrabal} show that the $^{16}$OH/$^{18}$OH opacity ratio derived from the radio lines is only a good measure of the actual abundance ratio in cold clouds with kinetic temperatures $\leq$20~K. This is due to excitation anomalies in the hyperfine levels caused by rotational pumping by the FIR lines. This certainly occurs at the velocity of \object{Sgr~B2} as can be seen from the higher energy lines of $^{16}$OH which show absorption within the $^{2}\Pi_{3/2}$ ladder but emission within the $^{2}\Pi_{1/2}$ ladder \citep{goicoechea_b}. This would cause the previously derived isotopic ratios to underestimate the true value. At the velocity of \object{Sgr~B2}, our fit gives a ratio of 320$^{+70}_{-30}$. This appears to confirm the suggestion of \citet{bujarrabal} that the ratio in the Galactic Centre has been underestimated (assuming the same excitation conditions exist for $^{16}$OH and $^{18}$OH and the ground state ratio accurately reflects the isotopic abundance). We also find a high value for the component attributed to the Galactic Bar ($-108$~km~s$^{-1}$) of 460$^{+90}_{-50}$. \citet{bujarrabal} find a ratio of $\simeq$370 in this component. 

In contrast, measurements of other molecules in \object{Sgr~B2} seem to agree with the lower radio OH values. However, none of these methods are easy to directly interpret and so may also underestimate the true value. Both H$_{2}$CO and CO must be used in a double ratio, relying on an accurate value of $^{12}$C/$^{13}$C. \citet{henkel} calculate $^{16}$O/$^{18}$O$\sim$200 from measurements of H$_{2}^{13}$C$^{16}$O/H$_{2}^{12}$C$^{18}$O, although the final value is uncertain due to the correction for radiation trapping in H$_{2}^{12}$C$^{16}$O (used to determine the associated $^{12}$C/$^{13}$C ratio). Observations of CO indicate a value of 250$\pm$100 in the Galactic Centre \citep{penzias} but this may be affected by chemical fractionation of $^{13}$C \citep{penzias_b}. \citet{gardner89} have observed CH$_{3}^{16}$OH and CH$_{3}^{18}$OH towards \object{Sgr~B2} leading to an isotopic ratio of 210$\pm$40. This does not include a correction for radiation trapping which would increase the ratio. \citet{nummelin_b} calculate even lower ratios from S$^{16}$O/S$^{18}$O (120$^{+58}_{-54}$) and S$^{16}$O$_{2}$/S$^{16}$O$^{18}$O (112$^{+26}_{-22}$). They explain these by assuming that the observed emission originates in the central regions of the \object{Sgr~B2} complex where $^{18}$O could be enhanced by the ejecta from massive stars. Our FIR absorption measurements trace gas in the outer parts of the \object{Sgr~B2} envelope.

Without low values at the Galactic Centre, our results show a much less convincing gradient than that obtained from the CO and H$_{2}$CO measurements. However, we have low number statistics and the distances of some components from the Galactic Centre are not very well known. There also must be some effect from a real variation in clouds at the same galactocentric distance, as indicated by the anomalous $^{16}$OH/$^{18}$OH and $^{12}$C/$^{13}$C at 0~km~s$^{-1}$. This means that a larger sample is necessary to accurately confirm the trend through the Galaxy.

\section{Comparison with other species \label{otherspecies}}

\begin{table}
\caption{Ratio of our derived $^{16}$OH column densities with those of CH from \citet{polehampton_ch} for each of the velocity components from Table~\ref{tab_col}.}
\label{tab_ratios}
\centering
\begin{tabular}{cc}
\hline 
\hline 
Velocity          & [$^{16}$OH/CH] \\
(km~s$^{-1}$)     &                \\
\hline  
$-108$            & $6.2\pm6.5$    \\
$-82$             & $35\pm21$      \\
$-52$/$-44.0$     & $19\pm15$      \\
$-24.4$           & $28\pm11$      \\
$+1.1$            & $38\pm18$      \\
$+15.7$           & $12\pm8$       \\
$+31$             & $36\pm35$      \\
$+52.8$/$+66.7$   & $34\pm7$       \\
\hline 
\end{tabular}
\end{table}

Several other related species have been observed in the line of sight clouds towards \object{Sgr~B2}. Observations of absorption due to the H$_{2}^{16}$O and H$_{2}^{18}$O ground state rotational lines have been made using SWAS \citep{neufeld_b}. Although H$_{2}^{16}$O completely absorbs the continuum, they calculate column densities in three line of sight velocity ranges from H$_{2}^{18}$O. Both OH and H$_{2}$O are formed by the dissociative recombination of H$_{3}$O$^{+}$ \citep[H$_{3}$O$^{+}$ is also observed to show absorption in the line of sight clouds;][]{goicoechea_c}. \citet{neufeld_c} show that if recombination of H$_{3}$O$^{+}$ is the only production mechanism, the ratio of OH to H$_{2}$O gives good constraints on the branching ratios. However, they also show that in warm gas, neutral-neutral reactions can alter the balance. A further effect is the inclusion of grain surface reactions, which can increase the gas-phase abundance of H$_{2}$O whilst OH is relatively unaffected \citep{plume}. 

Using the ortho-H$_{2}^{18}$O column densities from \citet{neufeld_b} with an ortho-to-para ratio of 3 and our $^{16}$O/$^{18}$O ratios averaged in their velocity ranges gives [H$_{2}$O/OH] ratios of 0.6--1.2 towards \object{Sgr~B2}. These values are a similar order of magnitude to those previously derived for line of sight absorption features towards \object{W51} \citep[$\sim$0.3; ][]{neufeld_c} and \object{W49} \citep[0.26, 0.43; ][]{plume}, indicating a high branching ratio for H$_{2}$O and/or the necessity of including other processes such as gas-grain interactions. In gas at the velocity of \object{Sgr~B2} itself, \citet{goicoechea_b} have derived even higher values of [H$_{2}$O/OH]$\sim$1-10. In this case, there is probably a warm gas component present and neutral-neutral reactions could be important.

In diffuse clouds, CH has been observed to be tightly correlated with OH, with a column density ratio, [OH/CH]$\sim$3 \citep[e.g.][]{liszt_ch}. Absorption from the ground state rotational transition of CH was observed as part of the ISO spectral survey towards \object{Sgr~B2} and has been fitted in a similar way to the OH lines \citep{polehampton_ch}. Table~\ref{tab_ratios} shows the ratio of our fitted $^{16}$OH column densities those of CH in each velocity component. These have a large uncertainty but seem to indicate values that could be up to ten times higher than that seen in diffuse clouds. This increased ratio cannot be caused by CH because the abundances found are very similar to diffuse cloud models \citep{polehampton_ch}. We can estimate the abundance of OH in the line of sight clouds using molecular hydrogen column densities in the range 4 to 14$\times10^{21}$~cm$^{-2}$ \citep[see][]{greaves96}. This leads to OH abundances 10$^{-7}$--10$^{-6}$ in these features. At the velocity of \object{Sgr~B2}, \citet{goicoechea_b} find an abundance of (2--5)$\times10^{-6}$. These abundances are much higher than found in diffuse clouds \citep[models of diffuse clouds predict $X$(OH)$\sim10^{-8}$;][]{vandishoeck86} and lead to the high [OH/CH] ratio. Such high abundances of OH can be found in PDR regions.

\section{Summary}

In this paper we have presented an analysis of $^{16}$OH and $^{18}$OH FIR rotational lines towards \object{Sgr~B2}. Absorption at different velocities along the line of sight allowed us to calculate the $^{16}$OH/$^{18}$OH ratio between the Sun and Galactic Centre.

The advantages of this technique for determining the isotopic ratio of oxygen are:
\begin{itemize}
\item Chemical fractionation effects are not important for $^{16}$OH and $^{18}$OH. Therefore the molecular column density ratio directly measures the isotope abundance.
\item There are optically thin lines present from both species.
\item The final ratio does not depend on the value of $^{12}$C/$^{13}$C as for previous H$_{2}$CO and CO measurements.
\item Uncertainties in the population of hyperfine states do not affect the final ratios (as they do for radio observations of $\Lambda$-doublet transitions).
\item Only ground state lines are seen for the line of sight clouds, showing that excitation effects are not important in these clouds.
\item All observations come from a single dataset with consistent and stable calibration over lines of both isotopologues.
\end{itemize}

The main disadvantage is the relatively low spectral resolution compared to the velocity separation of line of sight components. This introduced errors for individual components in the line shape modelling. However, the data used here allowed us to maximise confidence in the observed line shapes by combining several observations and with the good coverage of baselines around the lines.

We find  $^{16}$OH/$^{18}$OH ratios broadly consistent with previous isotopic abundances, although our results do not provide such clear evidence for a gradient of the ratio through the Galaxy (but they do not rule it out). In velocity components associated with the Galactic Centre, we find slightly higher ratios than previous results. This could be due to an underestimate in the previous radio observations of OH caused by pumping of hyperfine levels by the FIR rotational transitions \citep[see][]{bujarrabal}.

In comparing the total $^{16}$OH column densities with those of H$_{2}$O towards \object{Sgr~B2}, we find similar magnitude abundance ratios as have been seen in other lines of sight towards \object{W51} and \object{W49} \citep{neufeld_c,plume}. A comparison of $^{16}$OH with the column density of CH shows a wide scatter and large error, but appears to indicate higher ratios that previously observed in diffuse clouds.

This study shows that FIR rotational lines are an extremely useful tool for examining the variation of isotopic ratios in the ISM. The column densities of OH can be accurately derived as the method avoids the anomalous excitation that may be present in radio $\Lambda$-doublet lines. Ideally, future observations at higher spectral resolution are required to improve on the results. This may be possible using future telescopes such as SOFIA.

\begin{acknowledgements}

We would like to thank R. Laing, J. Hatchell, T. Wilson and F. van der Tak for helpful discussions, and the referee, C. Kahane, for several useful suggestions. The ISO Spectral Analysis Package (ISAP) is a joint development by the LWS and SWS Instrument Teams and Data Centres. Contributing institutes are CESR, IAS, IPAC, MPE, RAL and SRON. LIA is a joint development of the ISO-LWS Instrument Team at the Rutherford Appleton Laboratory (RAL, UK - the PI Institute) and the Infrared Processing and Analysis Center (IPAC/Caltech, USA). 

\end{acknowledgements}

\bibliographystyle{aa}
\bibliography{thesis_references}

\appendix

\section{Observations and wavelengths}

\begin{table*}
\caption{Log of the observations used.}
\label{tdts}
\centering
\begin{tabular}{lccccc}
\hline
\hline
Transition  & Observing & ISO    & LWS      & Spectral   & Repeated \\
            & Mode      & TDT    & Detector & Resolution & scans\\
            &           & Number &          & (km~s$^{-1}$) &\\
\hline 
\multicolumn{6}{l}{$^{16}$OH} \\
$^{2}\Pi_{3/2}~J$=5/2--3/2 (119~$\mu$m) & L03  & 50601013 & LW2 & 33 & 3\\
\hline 
$^{2}\Pi_{1/2}$--$^{2}\Pi_{3/2}$  & L03   & 50400823 & SW2 & 61 & 3\\
$J$=3/2--3/2 (53~$\mu$m)          & L03   & 50601013 & SW2 & 61 & 3\\
                                  & L03   & 50700707 & SW2 & 61 & 3 \\
                                  & L03   & 50800317 & SW2 & 61 & 3 \\
\hline 
$^{2}\Pi_{1/2}$--$^{2}\Pi_{3/2}$& L03   & 50900521 & SW5 & 40 & 3 \\
$J$=1/2--3/2 (79~$\mu$m)  & L03   & 50600814 & SW5 & 40 & 3 \\
 & L03   & 50600603 & SW4 & 40 & 3\\
 & L03   & 50700208 & SW4 & 40 & 3\\
 & L03   & 83800606 & SW4 & 40 & 4 \\
 & L03   & 50600506 & SW4 & 40 & 3 \\
\hline 
\multicolumn{6}{l}{$^{18}$OH} \\
$^{2}\Pi_{3/2}~J$=5/2--3/2 (120~$\mu$m) & L03 & 50601013 & LW2 & 33 & 3 \\
  & L03 & 50700610 & LW2 & 33 & 3 \\
 & L04 & 49401705 & LW2 & 33 & 15  \\
  & L04 & 46201118 & LW2 & 33 & 11 \\
  & L04 & 46900332 & LW2 & 33 & 14 \\
\hline
\end{tabular}
\end{table*}

\begin{table}
\caption{Wavelengths and Einstein coefficients ($A_{ij}$) for $^{16}$OH and $^{18}$OH averaged over hyperfine structure. The $^{16}$OH wavelengths are from \citet{brown} and \citet{varberg} and the Einstein coefficients were calculated from the line strengths of \citet{brown}. The $^{18}$OH wavelengths are from \citet{morino} with Einstein coefficients calculated from the line strengths of \citet{comben}.}
\label{wavelengths}
\begin{center}
\begin{tabular}{ccc}
\hline 
\hline
Transition         & Wavelength & $A_{ij}$    \\
                   &    ($\mu$m)& (s$^{-1}$)  \\
\hline
 \multicolumn{3}{l}{$^{16}$OH}                \\ 
$J$=3/2--3/2$^{a}$ & 53.2615   & 0.04559      \\
                   & 53.3512   & 0.04481      \\
$J$=1/2--3/2$^{a}$ & 79.1176   & 0.03575      \\
                   & 79.1812   & 0.03545      \\
\hline
 \multicolumn{3}{l}{$^{18}$OH}                \\
$J$=5/2--3/2       &119.9651   & 0.1364       \\
                   &120.1718   & 0.1357       \\
\hline  
\end{tabular}
\end{center}
$^{a}$ Cross ladder transition: $^{2}\Pi_{1/2}$--$^{2}\Pi_{3/2}$
\end{table}

\end{document}